\documentstyle[11pt,IAU207_pasp,twoside,psfig]{article}
\begin{document}
\title{The fragmentation of expanding shells:\\
cloud formation rates and masses}
 \author{Jan Palou\v s, Richard W\" unsch and So\v na Ehlerov\' a}
\affil{Astronomical Institute, Academy of Sciences of the Czech Republic,
Bo\v cn\' \i \ II 1401, 141 31 Prague 4, Czech Republic}
\begin{abstract}
 The gravitational instability of expanding shells is discussed. 
    Linear and nonlinear terms are included in an analytical solution 
    in the static and homogeneous medium. We discuss 
    the interaction of modes and give the time needed for fragmentation. 
    Masses of individual fragments are also estimated and   
    their formation 
    rates and the initial mass function are derived. Results of 
    simulations are compared to observation.
\end{abstract}

\section{Introduction}

The time evolution
of the gravitational collapse of perturbations in decelerating,
isothermal shocked layers has been examined numerically and analytically
using linearized equations by Elmegreen (1994).
In this paper, we continue with nonlinear analysis of the gravitational 
instability of spherically symmetric shells expanding into the stationary 
homogeneous medium.
We modify the approach adopted by Fuchs (1996)
who described the fragmentation of uniformly rotating self-gravitating 
disks. 
The inclusion of higher order terms helps to determine with better 
accuracy when, where and how quickly the fragmentation  happens.
Formation rates and the mass distribution function  of fragments are derived 
from simulations and compared to the observed mass spectrum of molecular 
clouds. 
 
\section{Hydrodynamical and Poisson equations on the surface of a thin shell}

Equations as derived by W\" unsch \& Palou\v s (2001) are

\begin{equation} 
\label{mc}
\frac{\partial m}{\partial t}+\left( \nabla, m\vec v \right) = A\  V \rho_0,
\end{equation}

\begin{equation} 
\frac{1}{A}\frac{d(m\vec v)}{dt}=-c^2\nabla\Sigma-\Sigma\nabla
\Phi,
\label{motion}
\end{equation}

\begin{equation} 
\label{poisson}
\Delta\Phi = 4\pi G\Sigma\delta(z)\ ,
\end{equation}
where $V$ is the expansion velocity of the shell, 
$m$ is mass in the area $A$ on the surface of the shell, 
and $\vec v$ denotes 
a two dimensional velocity of surface flows. 
$\Sigma $ is the surface density, $c$ is 
the constant isothermal sound speed inside the cold shell, $\Phi$ is the 
gravitational potential generated by the mass distribution in the shell, G is 
the gravitational constant and $\delta (z)$ is a delta function of the 
space 
coordinate $z$ perpendicular to the surface of the shell. 

\subsection{The linear solution}

The solution of linearized Eq. (1 - 3) may be written in the form of
exponential functions for the perturbations: 
$\Sigma_{1 - L}$ (perturbed surface density) and 
$\vec v \sim e^{i\omega t}$.
The shell starts to be gravitationally unstable when 
\begin{equation}\label{omega12}
\omega  = i\frac{3V}{R} - \sqrt{-\frac{V^2}{R^2} +
\frac{\eta^2 c^2}{R^2} - \frac{2\pi G\Sigma_0\eta}{R}}
\end{equation}
is purely imaginary and negative. $R$ denotes the radius and  
$\Sigma _0$ the unperturbed surface density of the shell.
$\omega $ is the angular frequency, in the case of instability $i\omega $ 
gives the perturbation growth rate,  and $\eta $ the wave number of 
perturbations.
The time evolution of $\Sigma _0$ and of $\Sigma _{1 - L}$, is shown in 
Fig. 1 for shell expanding in the homogeneous medium of the constant volume 
density 1 cm$^{-3}$ and average atomic weight 1.3. The total energy 
released is $10^{53}$ erg and $c$ = 1 km s$^{-1}$.   
\begin{figure}
\centerline{\hbox{
\psfig{figure=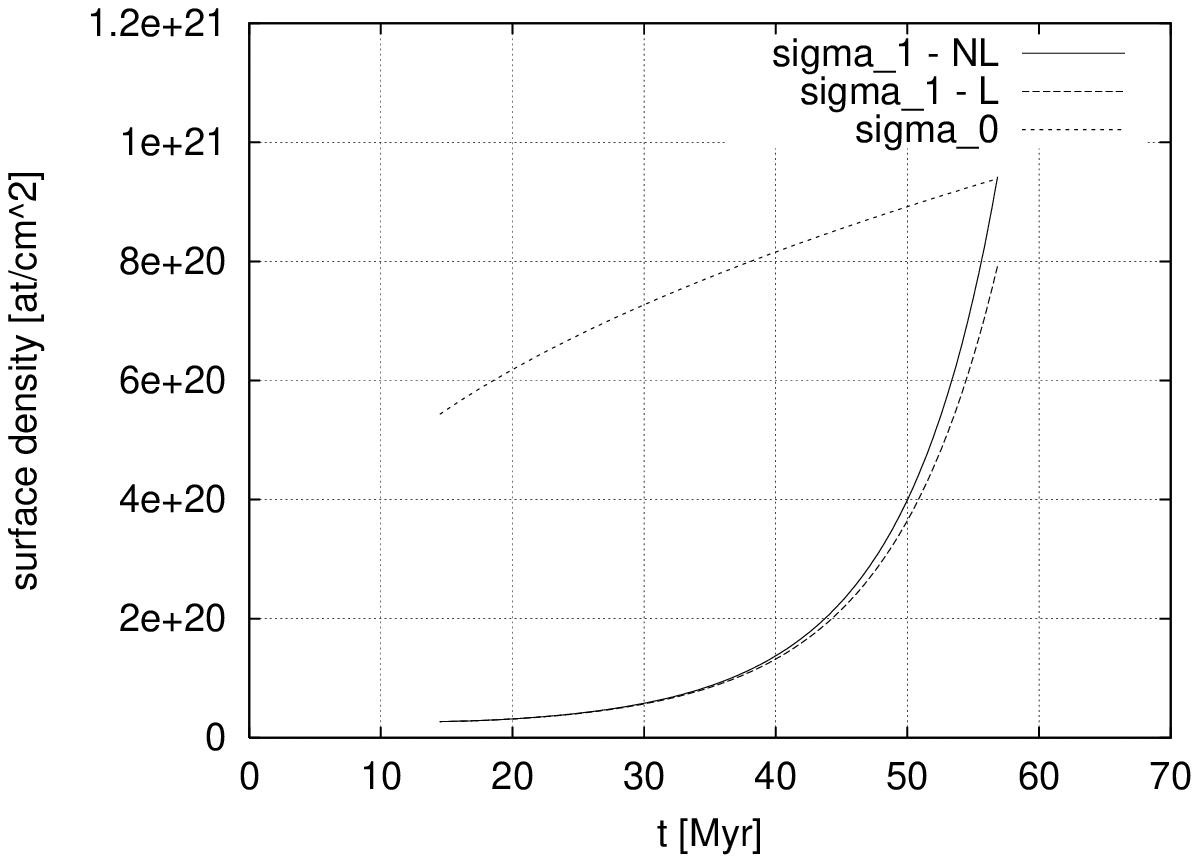,width=6.5cm,angle=0}
\psfig{figure=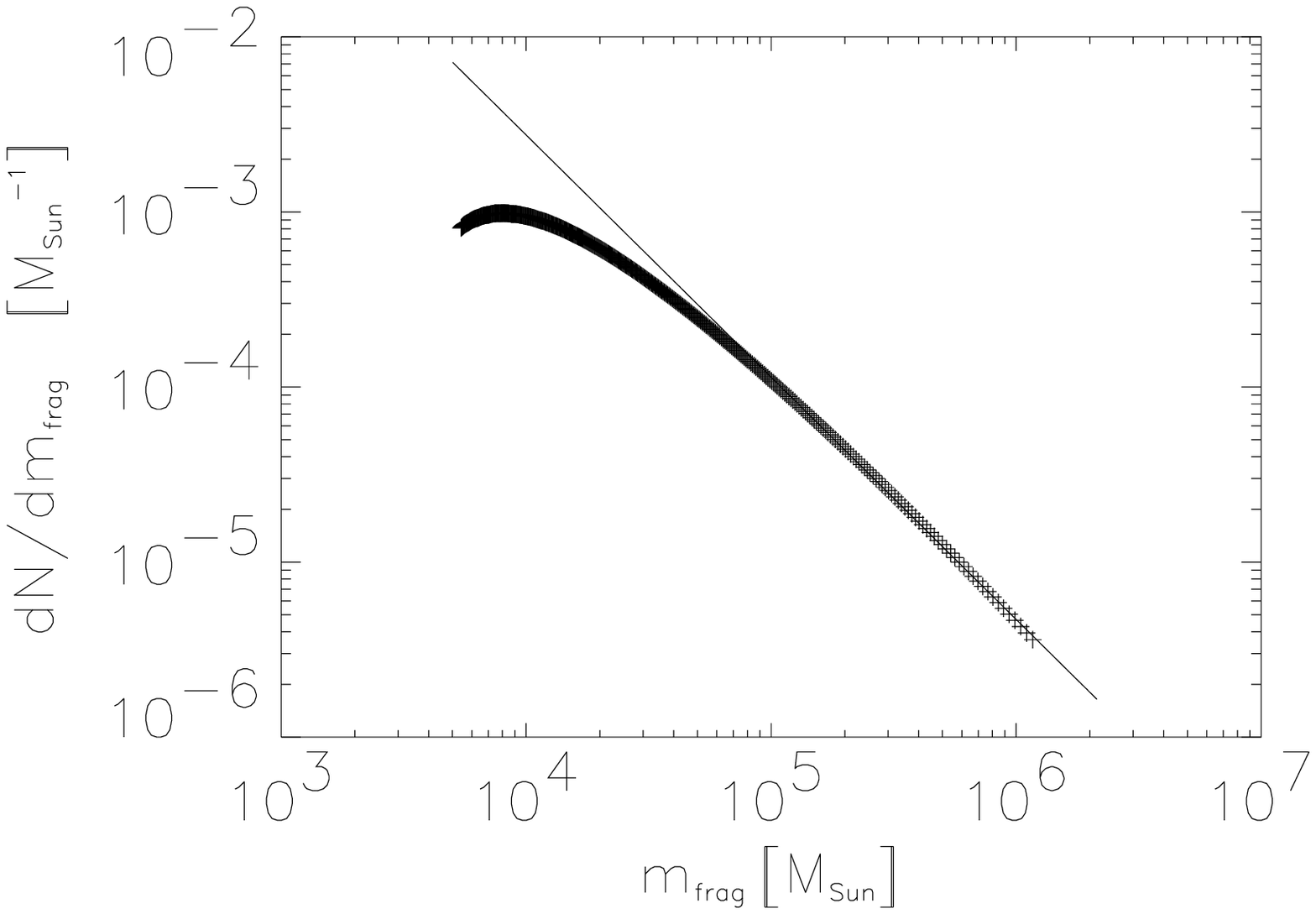,width=6.5cm,angle=0}
}}
\caption{Left: the evolution of the maximum surface density perturbation in the linear and
nonlinear case. Right: the mass spectrum of fragments. The straight line is the power law fit
of the decreasing part of the spectrum $m_{frag}^{-1.4}$.}
\end{figure}

\subsection{The nonlinear analysis}

W\" unsch \& Palou\v s (2001) include quadratic terms into equations 
for the evolution of perturbations. The time evolution of the
perturbed maximum surface density including nonlinear terms, $\Sigma_{1 - NL}$, 
is  given in Fig. 1. As another result of the nonlinear analysis 
Fig. 2 shows the spatial distribution of the surface density and velocity 
vectors of the surface flows resulting from mode interactions.  

\begin{figure}
\centerline{\hbox{
\psfig{figure=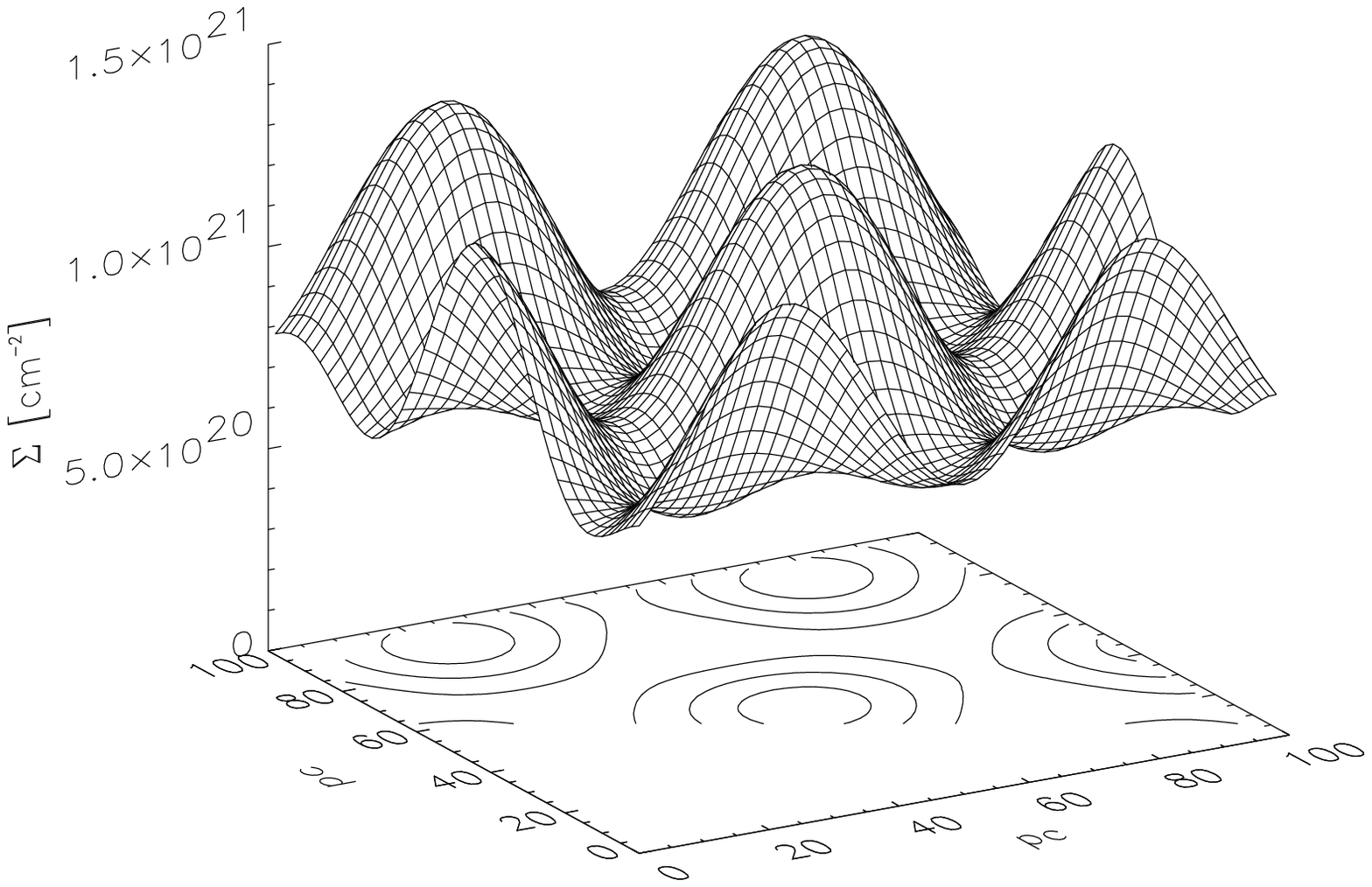,width=8cm,angle=0}
\psfig{figure=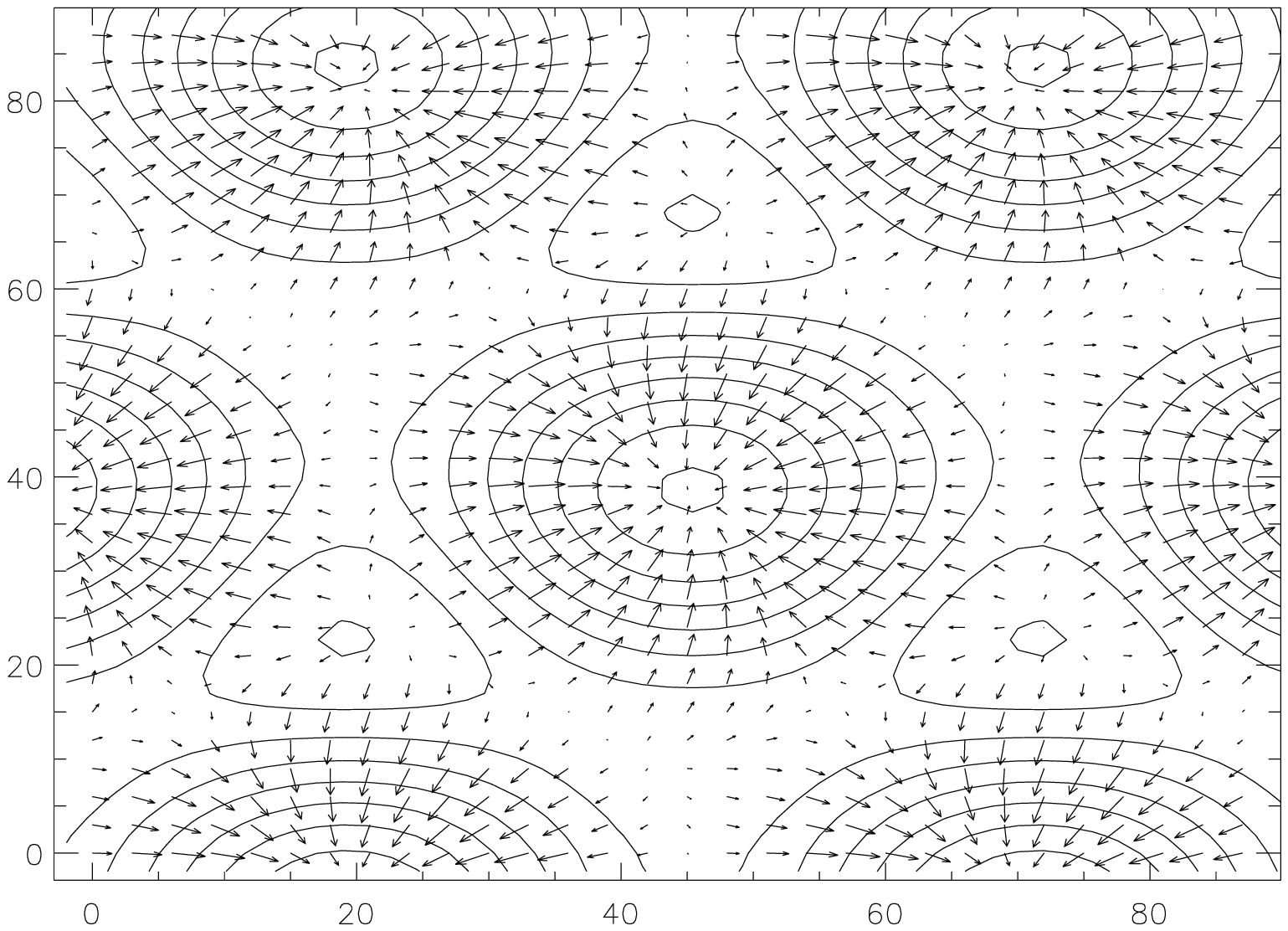,width=5cm,height=5cm,angle=0}
}}
\caption{Left: the spatial distribution of the surface density at $t=50$ Myr. 
Right: velocity
vectors of the surface flows at the same time.}
\end{figure}

\section{Time evolution of fragments and their mass spectrum}

After the time when the shell becomes unstable, the mass 
of the fragment decreases as its size decreases. 
Later, the inflow of mass begins to dominate (surface flows are
shown in Fig. 2) and the fragment mass grows. 
Masses of individual fragments are 
between a few times $10^3 M_{\odot}$ and $1.5 \times 10^6 M_{\odot }$ with
the highest frequency around $m_{frag} \sim 10^4 M_{\odot }$. 
Another consequence of the dispersion relation (4) is that the 
most massive fragments, corresponding to the smallest unstable wave number 
$\eta $, 
form very slowly. Therefore, the most massive clouds are created 
at the latest stages of the shell 
evolution. The mass spectrum of fragments is shown in Fig. 1. Its
decreasing part can be approximated with a power law
$dN/dm_{frag} \sim m^\alpha _{frag}$ where $\alpha = -1.4$. This is quite 
close to the observed mass spectrum of GMC in the Milky Way. Combes (1991)
gives $\alpha = -1.5$. NANTEN survey of the CO emission of the LMC 
(Fukui, 2001) gives steeper slope: $\alpha = -1.9$, which may be 
connected to higher 
level of random motions in the LMC compared to the Milky Way, which
restricts   the formation of late time massive fragments in the LMC 
and steepens their mass spectrum.

\vskip 1cm
{\it Acknowledgements} The authors gratefully acknowledge financial support 
by the Grant Agency of the Academy of Sciences of the Czech Republic under 
the grant No. A 3003705 and support by the grant project of the Academy of 
Sciences of the Czech Republic No. K1048102.

\end{document}